\documentclass{nature_withfigs}

\usepackage{amsmath,amsfonts,amssymb}
\usepackage{wrapfig}
\usepackage{graphicx}
\usepackage{bbm}
\usepackage{rotating}
\usepackage{ulem}

\title{Evidence for Efimov quantum states in an ultracold gas of cesium atoms}

\author{T. Kraemer$^{1}$, M. Mark$^{1}$,
P. Waldburger$^{1}$, J. G. Danzl$^{1}$, C. Chin$^{1,2}$, B.
Engeser$^{1}$, A. D. Lange$^{1}$, K.~Pilch$^{1}$, A. Jaakkola
$^{1}$, H.-C. N\"agerl$^{1}$ \& R. Grimm$^{1,3}$}

\begin{document}

\newcommand{\WW}{\ensuremath{\mathcal{W}}}
\newcommand{\QQ}{\ensuremath{\mathcal{Q}}}

\newcommand{\new}[1]{\textit{#1}}

\newcommand{\ketbra}[1]{\ensuremath{| #1 \rangle \langle #1 |}}
\newcommand{\ket}[1]{\ensuremath{|#1\rangle}}
\newcommand{\bra}[1]{\ensuremath{\langle#1|}}
\newcommand{\braket}[2]{\ensuremath{\langle #1|#2\rangle}}

\maketitle

\begin{affiliations}
\item Institut f{\"u}r Experimentalphysik, Universit{\"a}t
Innsbruck, Technikerstra{\ss}e 25, A--6020 Innsbruck, Austria
\item James Franck Institute, Physics Department of the University
of Chicago, 5640 S. Ellis Ave. Chicago, Illinois 60637, USA \item
Institut f\"ur Quantenoptik und Quanteninformation der
\"Osterreichischen Akademie der Wissenschaften,
Otto-Hittmair-Platz 1, A--6020 Innsbruck, Austria
\end{affiliations}

\begin{abstract}
Systems of three interacting particles are notorious for their
complex physical behavior. A landmark theoretical result in
few-body quantum physics is Efimov's
prediction\cite{Efimov70,Efimov71} of a universal set of bound
trimer states appearing for three identical bosons with a resonant
two-body interaction. Counterintuitively, these states even exist
in the absence of a corresponding two-body bound state. Since the
formulation of Efimov's problem in the context of nuclear physics
35 years ago, it has attracted great interest in many areas of
physics\cite{Jensen04,Lim77,Bruehl05,Braaten03,Braaten04,Stoll05}.
However, the observation of Efimov quantum states has remained an
elusive goal\cite{Jensen04,Bruehl05}. Here we report the
observation of an Efimov resonance in an ultracold gas of cesium
atoms. The resonance occurs in the range of large negative
two-body scattering lengths, arising from the coupling of three
free atoms to an Efimov trimer. Experimentally, we observe its
signature as a giant three-body recombination
loss\cite{Esry99,Braaten01} when the strength of the two-body
interaction is varied. We also detect a
minimum\cite{Nielsen99,Esry99,Bedaque00} in the recombination loss
for positive scattering lengths, indicating destructive
interference of decay pathways. Our results confirm central
theoretical predictions of Efimov physics and represent a starting
point with which to explore the universal properties of resonantly
interacting few-body systems\cite{Braaten04}. While Fesh\-bach
resonances\cite{Tiesinga93,Inouye98} have provided the key to
control quantum-mechanical interactions on the two-body level,
Efimov resonances connect ultracold matter\cite{ultracoldmatter}
to the world of few-body quantum phenomena.
\end{abstract}

Efimov's treatment of three identical
bosons\cite{Efimov71,Efimov70} is closely linked to the concept of
universality\cite{Braaten04} in systems with a resonant two-body
interaction, where the s-wave scattering length $a$ fully
characterizes the two-body physics. When $|a|$ greatly exceeds the
characteristic range $\ell$ of the two-body interaction potential,
details of the short-range interaction become irrelevant because
of the long-range
nature of the wave function. %($ |a| \gg \ell $).
Universality then leads to a generic behavior in three-body
physics, reflected in the energy spectrum of weakly bound Efimov
trimer states. Up to now, in spite of their great fundamental
importance, these states could not be observed experimentally. An
observation in the realm of nuclear physics, as originally
proposed by Efimov, is hampered by the presence of the Coulomb
interaction, and only two-neutron halo systems with a spinless
core are likely to feature Efimov states\cite{Jensen04}. In
molecular physics, the helium trimer is predicted to have an
excited state with Efimov character\cite{Lim77}. The existence of
this state could so far not be confirmed\cite{Bruehl05}. A
different approach to experimentally study the physics of Efimov
states is based on the unique properties of ultracold atomic
quantum gases. Such systems\cite{ultracoldmatter} facilitate an
unprecedented level of control to study interacting quantum
systems. The ultra-low collision energies allow to explore the
zero-energy quantum limit. Moreover, two-body interactions can be
precisely tuned based on Feshbach
resonances\cite{Tiesinga93,Inouye98}.

\begin{figure}
\begin{center}
\includegraphics[width=7cm]{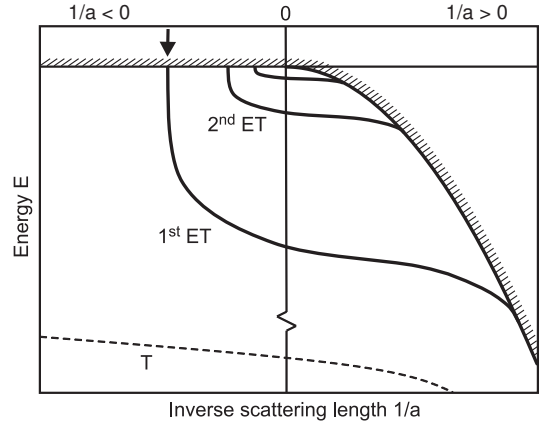}
\caption{Efimov's scenario: Appearance of an infinite series of
weakly bound Efimov trimer states (ET) for resonant two-body
interaction. The binding energy is plotted as a function of the
inverse two-body scattering length $1/a$. The shaded region
indicates the scattering continuum for three atoms ($a<0$) and for
an atom and a dimer ($a>0$). The arrow marks the intersection of
the first Efimov trimer with the three-atom threshold. To
illustrate the series of Efimov states, we have artificially
reduced the universal scaling factor from $22.7$ to $2$. For
comparison, the dashed line indicates a tightly bound non-Efimov
trimer\cite{Thomas35} (T) which does not interact with the
scattering continuum.} \label{Efimov}
\end{center}
\end{figure}

Efimov's scenario\cite{Efimov71,Efimov70,Braaten04} can be illustrated by the
energy spectrum of the three-body system as a function of the inverse
scattering length $1/a$ (Fig.~\ref{Efimov}). Let us first consider the
well-known weakly bound dimer state, which only exists for large positive $a$.
In the resonance regime, its binding energy is given by the universal
expression $E_b =-\hbar^2/(ma^2)$, where $m$ is the atomic mass and $\hbar$ is
Planck's constant divided by $2\pi$. In Fig.~\ref{Efimov}, where the resonance
limit corresponds to $1/a \rightarrow 0$, the dimer energy $E_b$ is represented
by a parabola for $a>0$. If we now add one more atom with zero energy, a
natural continuum threshold for the bound three-body system (shaded region in
Fig.~\ref{Efimov}) is given by the three-atom threshold ($E =0$) for negative
$a$ and by the dimer-atom threshold ($E_b$) for positive $a$. Energy states
below the continuum threshold are necessarily three-body bound states. When
$1/a$ approaches the resonance from the negative-$a$ side, a first Efimov
trimer state appears in a range where a weakly bound two-body state does not
exist. When passing through the resonance the state connects to the
positive-$a$ side, where it finally intersects with the dimer-atom threshold.
An infinite series of such Efimov states is found when scattering lengths and
energies are increased in powers of universal scaling
factors\cite{Efimov71,Efimov70,Braaten04} $e^{\pi/s_0} \approx 22.7$ and
$e^{2\pi/s_0} \approx 515$ ($s_0 = 1.00624$), respectively.

Resonant scattering phenomena arise as a natural consequence of Efimov's
scenario\cite{Efimov79}. When an Efimov state intersects with the continuum
threshold at negative scattering lengths $a$, three free atoms in the ultracold
limit resonantly couple to a trimer. This results in a triatomic Efimov
resonance. At finite collision energies, the phenomenon evolves into a
triatomic continuum resonance\cite{Bringas04}. Another type of Efimov
resonances\cite{Nielsen02} is found at positive values of $a$ for collisions
between a free atom and a dimer, when Efimov states intersect with the
dimer-atom threshold. While the latter type of Efimov resonances corresponds to
Feshbach resonances in collisions between atoms and dimers\cite{Nielsen02},
triatomic Efimov resonances can be interpreted as a three-body generalization
to Feshbach resonances\cite{Stoll05}.

Striking manifestations of Efimov physics have been predicted for
three-body recombination processes in ultracold gases with tunable
two-body
interactions\cite{Nielsen99,Esry99,Bedaque00,Braaten01,DIncao04,Braaten04}.
Three-body recombination leads to losses from a trapped gas with a
rate proportional to the third power of the atomic number density.
These losses are commonly described\cite{Weber03b} in terms of a
loss rate coefficient $L_3$. In the resonant case ($|a| \gg
\ell$), it is convenient to express this coefficient in the form
$L_3 = 3 \, C(a) \hbar a^4 /m$, separating a general
$a^4$-scaling\cite{Fedichev96,Weber03b} from an additional
dependence\cite{Esry99,Bedaque00,Braaten01} $C(a)$. Efimov physics
is reflected in a logarithmically periodic behavior $C(22.7a) =
C(a)$, corresponding to the scaling of the infinite series of
weakly bound trimer states. For negative scattering lengths, the
resonant coupling of three atoms to an Efimov state opens up fast
decay channels into deeply bound dimer states plus a free atom.
Triatomic Efimov resonances thus
show up in giant recombination loss. %features at $a = a_-$, $22.7a_-$, $...$.
This striking phenomenon was first identified in numerical
solutions to the adiabatic hyperspherical approximation of the
three-body Schr\"odinger equation assuming simple model potentials
and interpreted in terms of tunneling through a potential barrier
in the three-body entrance channel\cite{Esry99}. A different
theoretical approach\cite{Braaten01,Braaten04}, based on effective
field theory, provides the analytic expression $C(a) = 4590
\sinh{(2\eta_-)} / \left( \sin^2{[s_0 \ln{(|a|/a_-)} ]} +
\sinh^2{\eta_-} \right)$. The free parameter $a_-$ for the
resonance positions at $a_-$, $22.7\,a_-$, $...$
   depends on the short-range part of the three-body interaction and
is thus not determined in the frame of the universal long-range
theory. As a second free parameter, the dimensionless quantity
$\eta_-$ describes the unknown decay rate of Efimov states into
deeply bound dimer states plus a free atom and thus characterizes
the resonance width.

Our measurements are based on the magnetically tunable interaction
properties of Cs atoms\cite{Chin04} in the lowest internal state.
By applying fields between $0$ and $150$\,G, we varied the s-wave
scattering length $a$ in a range between $-2500\,a_0$ to
$1600\,a_0$, where $a_0$ is Bohr's radius. Accurate three-body
loss measurements are facilitated by the fact that inelastic
two-body loss is energetically forbidden\cite{Weber03b}. The
characteristic range of the two-body potential is given by the van
der Waals length\cite{vdWlength}, which for Cs is $\ell \approx
100\,a_0$. This leaves us with enough room to study the universal
regime requiring $|a| \gg \ell$. For negative $a$, a maximum value
of $25$ is attainable for $|a|/\ell$. Efimov's estimate
$\frac{1}{\pi}\,\ln(|a|/\ell)$ for the number of weakly bound
trimer states\cite{Efimov71} suggests the presence of one Efimov
resonance in the accessible range of negative scattering lengths.

\begin{figure}
\begin{center}
\includegraphics[width=9.8cm]{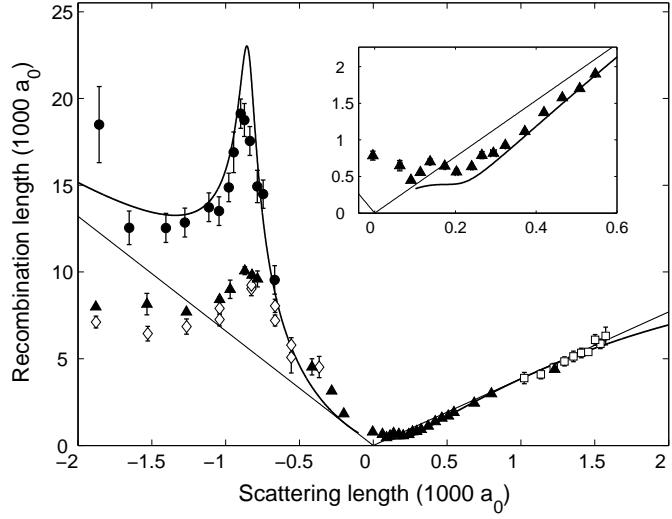}
\caption{Observation of the Efimov resonance in measurements of three-body
recombination. The recombination length $\rho_3 \propto L_3^{1/4}$ is plotted
as a function of the scattering length $a$. The dots and the filled triangles
show the experimental data from setup A for initial temperatures around 10\,nK
and 200\,nK, respectively. The open diamonds are from setup B at temperatures
of 250\,nK. The open squares are previous data\cite{Weber03b} at initial
temperatures between 250 and 450\,nK. The solid curve represents the analytic
model from effective field theory\cite{Braaten04} with $a_- = -850\,a_0$, $a_+
= 1060\,a_0$, and $\eta_- = \eta_+ = 0.06$. The straight lines result from
setting the $\sin^2$ and $\cos^2$-terms in the analytic theory to $1$, which
gives a lower recombination limit for $a<0$ and an upper limit for $a>0$. The
inset shows an expanded view for small positive scattering lengths with a
minimum for $C(a) \propto (\rho_3/a)^4$ near $210\,a_0$. The displayed error
bars refer to statistical uncertainties only. Uncertainties in the
determination of the atomic number densities may lead to additional calibration
errors for $\rho_3$ of up to 20\%.} \label{Rec_length}
\end{center}
\end{figure}

Our experimental results (Fig.~\ref{Rec_length}), obtained with
optically trapped thermal samples of Cs atoms in two different
set-ups (see methods), indeed show a giant loss feature marking
the expected resonance. We present our data in terms of a
recombination length\cite{Esry99} $\rho_3 =
[2m/(\sqrt{3}\hbar)\,L_3]^{1/4}$, which leads to the simple
relation $\rho_3/a = 1.36\,C^{1/4}$. Note that the general
$a^4$-scaling corresponds to a linear behavior in $\rho_3(a)$
(straight lines in Fig.~\ref{Rec_length}). A fit of the analytic
theory\cite{Braaten04,Braaten01} to our experimental data taken
for negative $a$ at temperatures $T\approx10$\,nK shows a
remarkable agreement and determines the resonance position to
$a_-=-850(20)\,a_0$ and the decay parameter to $\eta_-=0.06(1)$.
The pronounced resonance behavior with a small value for the decay
parameter ($\eta_- \ll 1$) demonstrates a sufficiently long
lifetime of Efimov trimers to allow their observation as distinct
quantum states.

All the results discussed so far are valid in the zero-energy
collision limit of sufficiently low temperatures. For ultralow but
non-zero temperatures the recombination length is unitarity
limited\cite{DIncao04} to $5.2 \, \hbar \, (m k_B T)^{-1/2}$.
%where the numerical factor was communicated to us by B.\ Esry and
%C.\ Greene.
For $T=10$\,nK this limit corresponds to about
$60,000\,a_0$ and our sample is thus cold enough to justify the
zero-temperature limit. For 250\,nK, however, unitarity limits the
recombination length to about $12,000\,a_0$. The Efimov resonance
is still visible at temperatures of 200 and 250\,nK (filled
triangles and open diamonds in Fig.~\ref{Rec_length}). The slight
shift to lower values of $|a|$ suggests the evolution of the
zero-energy Efimov resonance into a triatomic continuum
resonance\cite{Bringas04}. In further experiments at higher
temperatures (data not shown) we observed the resonance to
disappear above $\sim$500\,nK.

For positive scattering lengths, we found three-body losses to be typically
much weaker than for negative values. Our measurements are consistent with a
maximum recombination loss of $C(a)\approx 70$, or equivalently $\rho_3 \approx
3.9\,a$, as predicted by different theories\cite{Nielsen99,Esry99,Bedaque00}
(straight line for $a>0$ in Fig.~\ref{Rec_length}). For $a$ below $600\,a_0$
the measured recombination length significantly drops below this upper limit
(inset). The analytic expression from effective field
theory\cite{Bedaque00,Braaten04} for $a>0$ reads $C(a) = 67.1 \,e^{-2\eta_+} \,
\left( \cos^2[s_0\ln(a/a_+)] + \sinh^2\eta_+ \right) + 16.8\, (1-e^{-4\eta_+})$
with the two free parameters $a_+$ and $\eta_+$. The first term describes
recombination into the weakly bound dimer state with an oscillatory behavior
due to an interference effect between two different
pathways\cite{Nielsen99,Esry99}. The second term results from decay into deeply
bound states.
%with $a_+$ and
%$\eta$ being the two free parameters.
We use this expression to fit our data points with $a>5\,\ell
\approx 500a_0$. This somewhat arbitrary condition is introduced
as a reasonable choice to satisfy $a \gg \ell$ for the validity of
the universal theory. The fit is quite insensitive to the value of
the decay parameter and yields $\eta_+ < 0.2$. This result is
consistent with the theoretical assumption\cite{Braaten01} of the
same value for the decay parameter for positive and negative $a$,
which in our case is $\eta_+ = \eta_- =0.06$. For maximum loss, we
obtain $a_+ = 1060(70)\,a_0$. According to theory\cite{Braaten04},
the trimer state hits the dimer-atom threshold at $a = 1.1\,a_+
\approx 1170\,a_0$. The logarithmic periodicity of the Efimov
scenario suggests adjacent loss minima to occur at $\sqrt{22.7}
\times1060\,a_0 \approx 5000\,a_0$ and at $1060\,a_0/\sqrt{22.7}
\approx 220\,a_0$. While the former value is out of our accessible
range, the latter value ($a\approx 2\ell$) is too small to
strictly justify universal behavior in the resonance limit ($a \gg
\ell$). Nevertheless, our experimental results (inset to
Fig.~\ref{Rec_length}) indicate a minimum at $a \approx 210\,a_0$
and the analytic expression for $C(a)$ is found to describe our
data quite well down to this minimum.

\begin{figure}
\begin{center}
\includegraphics[width=7cm]{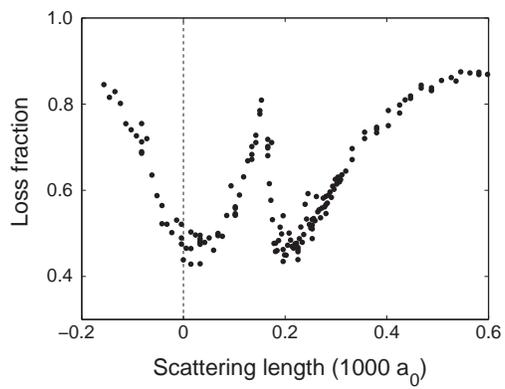}
\caption{Atom loss for small scattering lengths. Besides a minimum near zero
scattering length, we identify a minimum of recombination loss at
$\sim$$210\,a_0$, which can be attributed to a predicted destructive
interference effect\cite{Nielsen99,Esry99,Bedaque00}.} \label{loss_fraction}
\end{center}
\end{figure}

The occurrence of the interference minimum in three-body loss is
demonstrated more clearly in another set of experiments
(Fig.~\ref{loss_fraction}), where we simply measured the loss of
atoms after a fixed storage time in the optical trap. This minimum
is located at $a=210(10)\,a_0$ in addition to a second minimum
close to zero scattering length. We point out that the existence
of the minimum at $210\,a_0$ is very advantageous for efficient
evaporative cooling of Cs as it combines a large scattering cross
section with very low loss. Inadvertently, we have already
benefitted of this loss minimum for the optimized production of a
Bose-Einstein condensate of Cs\cite{Kraemer04}.

The comparison of our experimental results to available three-body
theory shows remarkable agreement, although the collision physics
of Cs is in general a very complicated multi-channel scattering
problem. We believe that the particular nature of the broad,
``open-channel dominated'' Feshbach resonance\cite{Koehler06} that
underlies the tunability of our system plays a crucial role. For
such a resonance, the two-body scattering problem can be reduced
to an effective single-channel model. It is a very interesting
question to what degree this great simplification of the two-body
physics extends to the three-body problem. Here we in particular
raise question how the regions of positive and negative scattering
lengths are connected in our experiment, where $a$ is changed
through a zero crossing, i.e.\ through a non-universal region, and
not across the universal resonance region. In our case, there is
no obvious connection between the Efimov state that leads to the
observed resonance for $a<0$ and the states responsible for the
behavior for $a>0$. In our analysis of the experimental data, we
have thus independently fitted the data sets for negative and
positive $a$. Nevertheless, the resulting values for the two
independent fit parameters $a_-$ and $a_+$ suggest a connection:
For the ratio $a_+/|a_-|$ our experiment yields 1.25(9), whereas
universal theory\cite{Braaten04} predicts 0.96(3). These numbers
are quite close in view of the Efimov factor of 22.7. If not an
accidental coincidence, we speculate that the apparent relation
between $a_+$ and $a_-$ may be a further consequence of
universality in a system where the resonant two-body interaction
can be modelled in terms of a single scattering channel. In
general, the multi-channel nature of three-body collisions near
Feshbach resonances\cite{Kartavtsev02,Petrov04} leads to further
interesting questions, like e.g.\ possible resonance effects
beyond the Efimov scenario. Advances in three-body theory are
necessary to finally answer these questions and to provide a
complete interpretation of our present observations.

In the past few years, applications of Feshbach resonances in ultracold gases
and the resulting possibility to create dimer states have set the stage for
many new developments in matter-wave quantum physics. The observation of an
Efimov resonance now confirms the existence of weakly bound trimer states and
opens up new vistas\cite{Braaten03,Stoll05} to experimentally explore the
intriguing physics of few-body quantum systems.

\bigskip
\noindent {\bf \large Methods}

\noindent {\bf Magnetic tuning of the two-body interaction}
\newline
For Cs atoms in their energetically lowest state (quantum numbers
$F=3$ for the total spin and $m_F=3$ for its projection) the
s-wave scattering length $a$ varies strongly with the magnetic
field\cite{Chin04}. Between 0 and 150\,G the dependence can in
general be well approximated by the fit formula
\begin{equation}
a(B)/a_0 = \left(1722+1.52\,B/{\rm G}\right) \left(1 -
\frac{28.72}{B/{\rm G}+11.74}\right), \nonumber
\end{equation}
except for a few narrow Feshbach resonances\cite{Chin04}. The
smooth variation of the scattering length in the low-field region
results from a broad Feshbach resonance centered at about $-12\,$G
(equivalent to $+12\,$G in the state $F=3$, $m_F=-3$).
%The
%character of this Feshbach resonance is open-channel
%dominated\cite{Koehler05,open}, i.e.\ two-body interactions can be
%modelled in terms of a single scattering channel or molecular
%potential.
In all our measurements we excluded the magnetic field
regions where the narrow Feshbach resonances influence the
scattering behavior through coupling to other molecular
potentials. The Efimov resonance is centered at 7.5\,G.

\noindent {\bf Trap setups and preparation of the Cs gases}
\newline All measurements were performed with
trapped thermal samples of Cs atoms at temperatures $T$ ranging
from 10 to 250\,nK. We used two different experimental setups,
which have been described elsewhere\cite{Kraemer04,Rychtarik04}.

In setup A we first produced an essentially pure Bose-Einstein
condensate (BEC) with up to 250,000 atoms in a far-detuned crossed
optical dipole trap generated by two 1060-nm Yb-doped fiber laser
beams\cite{Kraemer04}. We then ramped the magnetic field to
16.2\,G where the scattering length is negative with a value of
$-50\,a_0$, thus inducing a collapse of the
condensate\cite{Donley01}. After an equilibration time of 1\,s we
were left with a thermal sample at typically $T = 10$\,nK
containing up to 20,000 atoms at peak densities ranging from $n_0
= 3 \times 10^{11}$\,cm$^{-3}$ to $3 \times 10^{12}$\,cm$^{-3}$.
Alternatively, we interrupted the evaporation process before
condensation to produce thermal samples at $T \approx 200$\,nK in
a crossed dipole trap generated by one of the 1060-nm beams and a
$10.6$-$\mu$m CO$_2$-laser beam. After recompression of the trap
this produced typical densities of  $n_0 = 5 \times
10^{13}$\,cm$^{-3}$. The measurements in the region of the loss
minima as displayed in Fig.\,\ref{loss_fraction} were taken after
a storage time of 200\,ms at initial densities of $n_0 = 6 \times
10^{13}$\,cm$^{-3}$.

In setup B we used an optical surface trap\cite{Rychtarik04} in
which we prepared a thermal sample of 10,000 atoms at $T \approx
250$\,nK via forced evaporation at a density of $n_0 = 1.0 \times
10^{12}$\,cm$^{-3}$. The dipole trap was formed by a repulsive
evanescent laser wave on top of a horizontal glass prism in
combination with a single horizontally confining 1060-nm laser
beam propagating along the vertical direction.

\noindent {\bf Determination of three-body loss rate coefficients}
\newline
We measured three-body loss rates in setup A by recording the time
evolution of the atom number $N$ and the temperature $T$. A
detailed description of this procedure has been given in
Ref.~\cite{Weber03b}. In brief, the process of three-body
recombination not only leads to a loss of atoms, but also induces
``anti-evaporation'' and recombination heating. The first effect
is present at any value of the scattering length $a$. The second
effect occurs for positive values of $a$ when the recombination
products remain trapped. Atom loss and temperature rise are
modelled by a set of two coupled non-linear differential
equations. We used numerical solutions to this set of equations to
fit our experimental data. From these fits together with
measurements of the trapping parameters we obtained the rate
coefficient $L_3$. In setup B we recorded the loss at sufficiently
short decay times for which heating is negligible.

%% Here is the endmatter stuff: Supplementary Info, etc.
%% Use \item's to separate, default label is "Acknowledgements"

\begin{addendum}
%\item[Supplementary Information]

\item[Acknowledgements] We thank E. Braaten, C. Greene, B. Esry,
H. Hammer, and T. K{\"o}hler for many stimulating and fruitful
discussions and E. Kneringer for support regarding the data
analysis. We acknowledge support by the Austrian Science Fund
(FWF) within Spezialforschungsbereich 15 and within the Lise
Meitner program, and by the European Union in the frame of the TMR
networks ``Cold Molecules'' and ``FASTNet''. M.M. is supported
within the Doktorandenprogramm of the Austrian Academy of
Sciences.

\item[Competing Interests] The authors declare that they have no
competing financial interests.

\item[Correspondence] Correspondence and requests for materials
should be addressed to H.-C. N. (email:
christoph.naegerl@ultracold.at).

\end{addendum}

\newpage

\pagebreak

\newpage

\end{document}